\documentclass[galaxies, review, accept, pdftex, moreauthors]{Definitions/mdpi} 

\firstpage{1} 
\pubvolume{1}
\issuenum{1}
\articlenumber{0}
\pubyear{2024}
\copyrightyear{2024}
\externaleditor{Academic Editor: Ilaria Ruffa}
\datereceived{29 February 2024 } 
\daterevised{5 April 2024 } % Comment out if no revised date
\dateaccepted{12 April 2024 } 
\datepublished{ } 
\hreflink{https://doi.org/} % If needed use \linebreak
\usepackage{subcaption}
\Title{Investigating the Properties of the Relativistic Jet and Hot Corona in AGN with X-ray Polarimetry}

\TitleCitation{Investigating the Properties of the Relativistic Jet and Hot Corona in Agn with X-ray Polarimetry}

\Author{Dawoon E. Kim$^{1,2,3,}$*\orcidA{}, Laura Di Gesu $^{4}$\orcidB{}, Fr\'ed\'eric Marin $^{5}$\orcidC{}, Alan P. Marscher $^{6}$\orcidD{}, Giorgio Matt $^{7}$\orcidE{}, \mbox{Paolo Soffitta $^{1}$\orcidF{}}, Francesco Tombesi $^{3,8,9}$\orcidG{}, Enrico Costa $^{1}$\orcidH{} and Immacolata Donnarumma $^{4}$\orcidI{}}
\AuthorNames{Dawoon E. Kim, Laura Di Gesu, Fr\'ed\'eric Marin, Alan P. Marscher, Giorgio Matt, Paolo Soffitta, Francesco Tombesi, Enrico Costa and Immacolata Donnarumma}

\AuthorCitation{Kim, D.E.; Di Gesu, L.; Marin, F.; Marscher, A.P.; Matt, G.; Soffitta, P.; Tombesi, F.; Costa, E.; Donnarumma, I.}
\address{%
$^{1}$ \quad INAF Istituto di Astrofisica e Planetologia Spaziali, Via del Fosso del Cavaliere 100, 00133 Roma, Italy; paolo.soffitta@iaps.inaf.it (P.S.); enrico.costa@iaps.inaf.it (E.C.)\\ %MDPI: We added the email addresses here according to those submitted online at susy.mdpi.com. Please confirm.
$^{2}$ \quad Dipartimento di Fisica, Universit\`a degli Studi di Roma ``La Sapienza'', Piazzale Aldo Moro 5, 00185~Roma,~Italy\\
$^{3}$ \quad Dipartimento di Fisica, Universita\`a degli Studi di Roma ``Tor Vergata'', Via della Ricerca Scientifica 1, 00133~Roma,~Italy; francesco.tombesi@roma2.infn.it (F.T.)\\
$^{4}$ \quad ASI-Agenzia Spaziale Italiana, Via del Politecnico snc, 00133~Roma,~Italy; laura.digesu@asi.it (L.D.G.); immacolata.donnarumma@asi.it (I.D.)\\
$^{5}$ \quad Universit\'e de Strasbourg, CNRS, Observatoire Astronomique de Strasbourg, UMR 7550, 67000~Strasbourg,~France; frederic.marin@astro.unistra.fr (F.M.)\\ %MDPI: Address information should be sorted from subordinate to superior, please modify.    
$^{6}$ \quad Institute for Astrophysical Research, Boston University, 725 Commonwealth Avenue, Boston, MA 02215, USA; marscher@bu.edu (A.P.M.)\\
$^{7}$ \quad Dipartimento di Matematica e Fisica, Universit\`a degli Studi Roma Tre, Via della Vasca Navale 84, 00146~Roma,~Italy; giorgio.matt@uniroma3.it (G.M.)\\
$^{8}$ \quad Istituto Nazionale di Fisica Nucleare, Sezione di Roma ``Tor Vergata'', Via della Ricerca Scientifica 1, 00133~Roma,~Italy\\
$^{9}$ \quad Department of Astronomy, University of Maryland, College Park, MD 20742, USA\\
}

\corres{Correspondence: dawoon.kim@inaf.it}

\abstract{X-ray polarimetry has been suggested as a prominent tool for investigating the geometrical and physical properties of the emissions from active galactic nuclei (AGN). The successful launch of the Imaging X-ray Polarimetry Explorer (IXPE) on 9 December 2021 has expanded the previously restricted scope of polarimetry into the X-ray domain, enabling X-ray polarimetric studies of AGN. Over a span of two years, IXPE has observed various AGN populations, including blazars and radio-quiet AGN. In this paper, we summarize the remarkable discoveries achieved thanks to the opening of the new window of X-ray polarimetry of AGN through IXPE observations. We will delve into two primary areas of interest: first, the magnetic field geometry and particle acceleration mechanisms in the jets of radio-loud AGN, such as blazars, where the relativistic acceleration process dominates the spectral energy distribution; and second, the geometry of the hot corona in radio-quiet AGN. Thus far, the IXPE results from blazars favor the energy-stratified shock acceleration model, and they provide evidence of helical magnetic fields inside the jet. Concerning the corona geometry, the IXPE results {are consistent with} a disk-originated slab-like or wedge-like shape, as could result from Comptonization around the accretion disk.
}

\keyword{X-ray polarimetry; active galactic nuclei; relativistic jets}

\newcommand{\ixpe}{{IXPE }}

\newcommand{\pd}{{$\Pi$}}
\newcommand{\pa}{{$\psi$}}
\newcommand{\pdx}{{$\Pi_{\rm X}$}}
\newcommand{\pax}{{$\psi_{\rm X}$}}

\newcommand{\pdradi}{{$\Pi_{\rm R}$}}

\newcommand{\pdo}{{$\Pi_{\rm O}$}}
\newcommand{\pao}{{$\psi_{\rm O}$}}

\newcommand{\xmm}{{XMM-Newton}}
\newcommand{\nustar}{{NuSTAR}}
\newcommand{\swift}{{Swift}-XRT}

\def \ergs{\hbox{erg s$^{-1}$}}

\def \ergsc{\hbox{erg s$^{-1}$ cm$^{-2}$}}

\newcommand\aap{Astron. Astrophys.}                % Astronomy and Astrophysics
\newcommand\aj{Astron. J.}                   % Astronomical Journal (the)
\newcommand\apjs{Astrophys. J. Suppl.}               % Astrophysical Journal, Supplement
\newcommand\araa{Annu. Rev. Astron. Astrophys.}             % Annual Review of Astronomy and Astrophysics
\newcommand\mnras{Mon. Not. R. Astron. Soc.}             % Monthly Notices of the Royal Astronomical Society
\newcommand\na{New~Astron.}          % New Astronomy
\newcommand\prl{Phys. Rev.~Lett.}    % Physical Review Letters
\newcommand\pasp{Publ. Astron. Soc. Pac.}               % Publications of the Astronomical Society of the Pacific
\begin{document}

%%%%%%%%%%%%%%%%%%%%%%%%%%%%%%%%%%%%%%%%%%
%\setcounter{section}{-1} %% Remove this when starting to work on the template.
%\section{How to Use this Template}
%
%The template details the sections that can be used in a manuscript. Note that the order and names of article sections may differ from the requirements of the journal (e.g., the positioning of the Materials and Methods section). Please check the instructions on the authors' page of the journal to verify the correct order and names. For any questions, please contact the editorial office of the journal or support@mdpi.com. For LaTeX-related questions please contact latex@mdpi.com.%\endnote{This is an endnote.} % To use endnotes, please un-comment \printendnotes below (before References). Only journal Laws uses \footnote.

% The order of the section titles is different for some journals. Please refer to the "Instructions for Authors” on the journal homepage.

\section{Introduction}

Active galactic nuclei (AGN) stand out as fascinating and powerful sources (with bolometric luminosities $\sim$$10^{42}$--$10^{46}$ \ergs; {e.g.},~\cite{2002ApJ...579..530W, 2017A&A...603A..99P}) that emit electromagnetic radiation across the entire spectrum, ranging from radio {wavelengths} to gamma rays. {Specifically, in the X-ray regime, emissions originate from the vicinity of the central engine, in either the inner part of the jet or in an X-ray corona.} The high penetration power of X-ray radiation facilitates the unobstructed study of processes near black holes. {So far, X-ray probes have been limited to spectral and timing techniques \cite{2022hxga.book.....B}. Although these methods provide clues about the radiative processes producing X-rays in the inner AGN region, they are insensitive to some key characteristics, such as the geometry of the emitting region or the magnetic field, which so far have remained largely unconstrained.}

In this context, polarimetric studies have been suggested as a prominent tool for investigating the geometrical and physical properties of AGN emissions \cite{1979rpa..book.....R}. The successful launch of the Imaging X-ray Polarimetry Explorer (IXPE) on 9 December 2021 marked a significant advancement in polarimetry by extending it into the X-ray range \cite{weisskopf22}. {IXPE is a collaborative mission between NASA and the Italian Space Agency (Agenzia Spaziale Italiana, ASI). {It comprises three identical X-ray telescope systems, each one composed of one mirror module assembly and one detector unit hosting a gas pixel detector~\cite{2001Natur.411..662C} with a full calibration and filtering system. IXPE is designed to measure the linear polarization in the 2--8 keV band using polarization mapping techniques. It offers a field of view greater than $11'$ and an angular resolution of less than or equal to $30''$.} Over the first two years of operation, IXPE has detected X-ray polarization from several AGN, {helping to address the long-standing uncertainty about the jet magnetic field structure and acceleration process and the geometry of the X-ray corona.} 

Traditionally, AGN are divided into two main classes: radio-loud (RLAGN) and radio-quiet (RQAGN), based on their radio to optical flux ratio \cite{1995PASP..107..803U, 2017A&ARv..25....2P}. {In some cases, they are classified as jetted or non-jetted based on the presence or absence of strong {relativistic jets}~\cite{2017NatAs...1E.194P}.} Among the population of RLAGN, blazars are ideal targets for studying relativistic jets because, in these sources, {the jet flow is relatively closely aligned} to the observer's line of sight, and their emission is {dominated by jet emission} across the entire spectral energy distribution (SED) due to relativistic Doppler boosting, e.g., \cite{Hovatta2016}. The SED of blazars  exhibits two broad non-thermal radiation components as shown in Figure \ref{fig:spectra}. One component, located in the lower-frequency range, covers emissions from radio to optical/UV (and, in some cases, to X-ray) and is generally ascribed to synchrotron emission from relativistic electrons. The other component may be explained in two different scenarios: leptonic or hadronic models. In the leptonic process, the high-energy emission is attributed to Compton scattering of synchrotron photons (synchrotron self-Compton or SSC; e.g., \citep{1974ApJ...188..353J,1992ApJ...397L...5M}) or photons {originating} outside the jet (external Compton or EC; e.g., \citep{Dermer1993, Sikora1994,1987ApJ...322..650B}). In the hadronic scenario, proton synchrotron emission, $\pi_0$ decay photons, synchrotron, and Compton emission from secondary decay products of charged pions have been suggested, e.g., \citep{1993A&A...269...67M, 2000NewA....5..377A, Bottcher2013}. These emission processes are predicted to produce different X-ray polarization properties depending on the acceleration mechanism in both leptonic and hadronic scenarios \cite{2019ARA&A..57..467B}. Furthermore, depending on the position of the peak frequency of the electron synchrotron emission hump ($\nu_{peak}$), {blazars} are separated into low- ($\nu_{peak} < 10^{14}~$Hz or 3 $\upmu$m; LSP), intermediate- ($\nu_{peak}$ $\sim 10^{14}~$Hz--$10^{15}~$Hz; ISP), and high-synchrotron-peak ($\nu_{peak} > 10^{15}~$Hz or {0.3} $\upmu$m; HSP) subclasses \citep{2010ApJ...716...30A}. Therefore, studying various subclasses of blazars with X-ray polarimetry allows us to explore different physical processes. First, with HSP targets, the particle acceleration process and geometrical features of the magnetic field inside the jets can be studied. Second, with ISP and LSP blazars, the superposition of synchrotron emission and leptonic or hadronic emission in spectra enables tests of the acceleration mechanisms behind the high-energy emission hump.

In the case of RQAGN, given that they are not {typically} dominated by jet emissions, we can observe other emission components surrounding the central engine (see  Figure \ref{fig:spectra}). The primary X-ray emission of AGN is believed to originate from a corona of hot (with a temperature of 10--100 keV) and optically thin (or mildly thick) plasma {that Compton scatters} the optical/UV photons from the accretion disk, e.g., \cite{1991ApJ...380L..51H}. {Additionally, these X-ray photons are reflected (backscattered) either by the accretion disk or by the {more} distant torus, which results in additional X-ray spectral components as secondary emissions}~\cite{2018MNRAS.478..950M}. RQAGN are divided into Seyfert 1 and 2 galaxies based on {whether or not both broad and narrow emission lines are observed} in the optical band, which partially depends on the differences in the line of sight direction \cite{1993ARA&A..31..473A}. In the X-ray regime, Seyfert 1 galaxies, which have a relatively small inclination angle between our line of sight and the axis of the system, {averaging} $39^{\circ}$ \cite{2001ApJ...561L..59W, 2022A&A...659A.123S}, show unobscured X-ray coronae and disk reflection, while the emission from Seyfert 2s is {dominated by photons reprocessed} through reflection from the {strongly absorbing} dusty torus and from the polar winds due to the larger inclination angle.

The spectrum of the corona typically follows a power law with a high-energy cutoff, e.g., \citep{1980A&A....86..121S}. The slope of the power law is determined by the electron temperature and the optical depth, while the cutoff energy depends mostly on the electron temperature. The size and geometry of the corona remain elusive, {although} these characteristics are crucial for differentiating between models for the formation of AGN coronae, e.g., \citep{1998MNRAS.299L..15D}. Regarding the geometric characteristics of the hot corona, several possibilities have been suggested, including lamppost, spherical, conical, slab, and wedge. In particular, spherical or conical structures located on the rotation axis above and below the black hole would suggest a potential connection between the corona and the jet, e.g., \citep{2004A&A...413..535G, 2005ApJ...635.1203M}. In this context, it is predicted that the X-ray polarization of the corona emission depends on the geometry, e.g., \citep{1993MNRAS.261..346H}, and demonstrates that X-ray polarization measurements can serve as a powerful tool for probing the disk/corona geometry, e.g., \citep{1989ESASP.296..991M, 1993MNRAS.264..839M}. Hence, the polarization observations of unobscured RQAGN {can} provide valuable insights into the geometric features of the~corona.

In this short review, we will briefly summarize the theoretical expectations and remarkable discoveries obtained from the first two years of \ixpe observations of RLAGN and RQAGN. These concern the magnetic field geometry and particle acceleration mechanisms in the jets, as well as the geometry of the hot corona. We will then discuss future~prospects.

\begin{figure}[H]
       
         \includegraphics[width=0.6\textwidth]{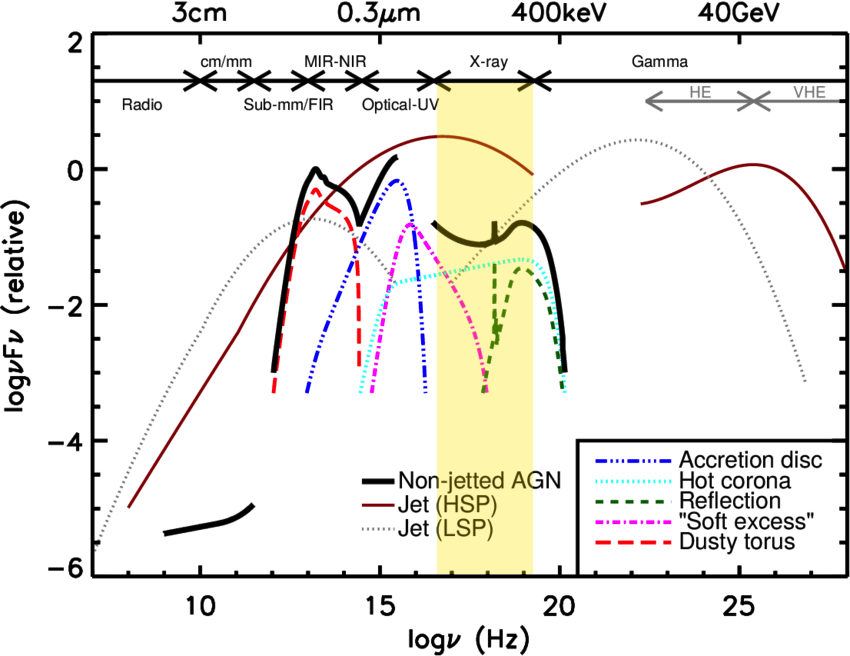}
\caption{A schematic representation of {SED for RLAGN (HSP: red solid line, and LSP: gray dotted line) and RQAGN (black solid line with labeled colored lines for different emission components). The yellow shaded area indicates the X-ray regime.} The figure is reproduced from {Figure 1} in \cite{2017A&ARv..25....2P}.}\label{fig:spectra}
\end{figure}

%\begin{figure}[t]
%\centering         
%\includegraphics[width=0.75\columnwidth]{figure/agn_spectra}
%\caption{A schematic representation of an AGN SED. Image credit: C. M. Harrison. Figure reproduced from \cite{2017FrASS...4...35P}\label{agn_spectrra}}
%\end{figure}   
%\unskip

%%%%%%%%%%%%%%%%%%%%%%%%%%%%%%%%%%%%%%%%%%
\section{X-ray Polarization from Relativistic Jets \label{sec:xpol_jet}}

\subsection{Theoretical Expectations}

{The observed emission coming from relativistic jets is often well-described by a power law spectrum within a given frequency band. Theoretically, this can be explained as synchrotron radiation produced by a power law energy spectrum of electrons with Lorentz factor} $\gamma$, $N_e(\gamma) \propto \gamma^{-p}$, which produces a power law synchrotron radiation spectrum $F_{\nu} \propto \nu^{-\alpha}$, with $\alpha = (p-1)/2$. This results in a polarization degree (\pd) of \cite{1979rpa..book.....R} 
\begin{equation} \label{eq:1}
	\Pi_{Sync} = \frac{p+1}{p+7/3} = \frac{\alpha+1}{\alpha+5/3}.
\end{equation}

%MDPI: We indented this, please confirm, The following highlights are the same.
Considering the typical power law index obtained from blazars {$\sim$$p$ = 2--3}, a polarization degree of the electron synchrotron emission in a homogeneous magnetic field is expected to be $\Pi$$\sim$$70\%$. The polarization angle (\pa) is perpendicular to the averaged magnetic field direction projected onto the sky plane. {However, in reality, polarization can be reduced by the geometrical and physical properties within the jet, such as temporal variation effects occurring over timescales shorter than the integration time of the polarization measurement}~\cite{2005AJ....130.1418J, 2017MNRAS.472.3589K,2018MNRAS.474.1296B}. Thereby, in order to explain the polarization measurements and variability, various geometrical and physical models have been suggested. For instance, for the geometrical model, there have been several suggestions: a bending jet \cite{2013ApJ...768...40L}, a spiral motion model in which a blob propagates in a spiral trajectory within a helical magnetic field~\cite{Marscher2008}, and the ballistic motion of an emission blob in a jet with a helical magnetic field \cite{2017MNRAS.467.3876L}. In the case of the physical conditions, there have been several suggestions as well: many turbulent cells, {the TEMZ model,} \citep{2014ApJ...780...87M,2017Galax...5...63M, 2021Galax...9...27M}, shock and kink instabilities that can efficiently dissipate jet bulk energy to accelerate particles \cite{2001MNRAS.328..393A, 2008ApJ...682L...5S, 2009ApJ...700..684M, 2018PhRvL.121x5101A}, and magnetic reconnection \cite{2014ApJ...783L..21S, 2014PhRvL.113o5005G, 2018ApJ...862L..25Z, 2022ApJ...924...90Z}. Refs.~\cite{2021Galax...9...37T, 2022A&A...662A..83D} have presented the predicted X-ray and simultaneous multiwavelength polarization characteristics among various theoretical models as follows: {(1) in the shock acceleration model, particle acceleration is characterized by a strong self-generated magnetic field. Hence, a large degree of polarization, \pdx$\sim$40$\%$, and relatively stable \pdx\ and \pax\ over time (variability in a few days to a week) are {predicted} \cite{2020MNRAS.498..599T, 2018MNRAS.480.2872T}. Moreover, a chromatic feature in X-ray and optical polarization, where \pdx/\pdo~$\gtrsim$~2, is anticipated because particles emitting at longer wavelengths, due to their longer synchrotron cooling time, travel away from the shock front to regions with a more disordered magnetic field; (2) in the magnetic reconnection model, the reconnection episodes are triggered by instability induced, for example, by current {sheets} \cite{2015ApJ...806..167G, 2014ApJ...783L..21S}. \pdx~$\lesssim$~$20-30 \%$ and similar characteristics for \pdo\ are predicted as a result of the simultaneous contribution of several active current sheets with different orientations. Additionally, the evolution of the instability results in significant variations, and \pax\ can show variability on a daily timescale; (3) in the TEMZ model, \pdx~$\lesssim$~$30 \%$ and rapid changes in \pdx\ and \pax\ less than a day timescale are predicted because of the superposition of the emission of several cells, in which the magnetic field is uniform but randomly oriented \cite{2014ApJ...780...87M}. In addition, mild chromatic behavior, \pdx/\pdo~$\lesssim$~2, is expected as a large number of cells can radiate at lower frequencies. Thus, the variability is erratic and somewhat driven by the dimension and number of cells. Consequently, expanding multiwavelength polarimetric measurements on synchrotron emission to the X-ray band provides clues about the layout of the magnetic field as particles emitting at different energies can assist in identifying the predominant acceleration process.}

Compton scattering of unpolarized seed photons by highly relativistic electrons produces {zero polarization in the scattered radiation} \cite{2013ApJ...774...18Z}. Therefore, detecting X-ray polarization from LSP blazars, whose emission component is primarily dominated by Compton scattering in X-rays, would effectively rule out the possibility of EC processes. In contrast, under the conditions of proton synchrotron radiation, the polarization levels could potentially reach around $70\%$, consistent with the proton spectral indices in Equation \eqref{eq:1}, where $p$ is approximately 2–3 \cite{2013ApJ...774...18Z}. A polarization degree significantly higher than that at mm-waves and similar to or greater than that at optical wavelengths, and a slope of the X-ray emission that is sufficiently flat, could be interpreted as an indication of proton synchrotron emission. Furthermore, for cases involving a combination of synchrotron emission at soft X-ray energies and SSC/EC at harder X-rays, detailed SED modeling and decomposition of the polarization properties of each component are required. The model considering multiple emission zones with turbulent magnetic fields predicts $\Pi_{SSC} / \Pi_{Sync} \approx 0.3$ \cite{2019ApJ...885...76P}.

\subsection{\ixpe Observations of High Synchrotron Peak Blazars}

\ixpe observations of HSP blazars have been conducted for Mrk 501 \cite{2022Natur.611..677L},\linebreak   Mrk 421 \mbox{\cite{2022ApJ...938L...7D, DiGesu2023, 2024A&A...681A..12K}}, 1ES1959+650 \cite{2024arXiv240104420E}, PG1553+113 \cite{2023ApJ...953L..28M},  and 1ES0229+200 \cite{2023ApJ...959...61E}. The X-ray polarization of each target was estimated through time-averaged analysis by considering the entire observation period and time- and energy-resolved analyses by dividing the observations into sub-segments using model independent event Stokes parameter analysis \cite{2015APh....68...45K} and spectropolarimetric analysis \cite{2017ApJ...838...72S}. In addition, during all the \ixpe blazar observations, a simultaneous multiwavelength polarimetry campaign was conducted, including radio, infrared (IR), and optical wavelengths, to examine the synchrotron emission properties across different wavelengths. In this section, we summarize the reported X-ray and multiwavelength polarimetry features of each observation, as well as the X-ray spectral properties. Table \ref{tab:xpol_hsp} presents a summary of the time-averaged multiwavelength polarization {information} for all the HSP observations, and Table \ref{tab:hsp} provides a summary of the X-ray spectral characteristics, including the photon index and flux derived based on the log parabolic model \cite{2004A&A...413..489M}. The log parabolic model represents power laws with a photon index that varies according to a log parabola in energy:
\begin{equation} \label{eq:logpar}
\begin{split}
N(E) = K(E/E_{pivot})^{(\alpha - \beta \log(E/E_{pivot}))},
\end{split}
\end{equation}
\noindent where the pivot energy $E_{pivot}$ is a scaling factor, $\alpha$ describes the slope of the photon spectrum at $E_{pivot}$, $\beta$ expresses the spectral curvature, and $K$ denotes a normalization constant.

\begin{table}[H] 
\caption{Contemporaneous multiwavelength polarization properties of HSPs. \label{tab:xpol_hsp}}
\newcolumntype{C}{>{\centering\arraybackslash}X}
\begin{tabularx}{\textwidth}{lCCC}
\toprule
\textbf{Source}	&  \textbf{X-ray}	& \textbf{Optical \& IR} \textsuperscript{a} & \textbf{Radio} \textsuperscript{a}\\
&\textbf{\boldmath{$\Pi(\%)$ \quad $\psi(^\circ)$}} & \textbf{\boldmath{$\Pi(\%)$ \quad $\psi(^\circ)$}} & \textbf{\boldmath{$\Pi(\%)$ \quad $\psi(^\circ)$}}\\
\midrule
Mrk~501 I   	\textsuperscript{1} & 10 $\pm$ 2 	\quad	134 $\pm$ 5	& 4 $\pm$ 1 	 	\quad	119 $\pm$ 9 	& 1.5 $\pm$ 0.5 	\;	 	152 $\pm$ 10 	\\
Mrk~501 II  	\textsuperscript{1} & 11 $\pm$ 2 	\quad 	115 $\pm$ 4	& 5 $\pm$ 1 		\quad 	117 $\pm$ 3	& -- 			\qquad	--			\\
Mrk~421 I   	\textsuperscript{2} & 15 $\pm$ 2 	\quad	35 $\pm$ 4	& 2.9 $\pm$ 0.5 	\quad	32 $\pm$ 5	& 3.4 $\pm$ 0.4 	\quad	55 $\pm$ 2	\\
Mrk~421 II  	\textsuperscript{3} & 10 $\pm$ 1 	\quad	Rotation	& 4.4 $\pm$ 0.4 	\quad	140 $\pm$ 6  	& 2.4 $\pm$ 0.1 	\quad	139 $\pm$ 8	\\
Mrk~421 III 	\textsuperscript{3} & 10 $\pm$ 1 	\quad 	Rotation	& 5.4 $\pm$ 0.4 	\quad	145 $\pm$ 1	& --			\quad	--			\\
Mrk~421 IV  	\textsuperscript{4} & 14 $\pm$ 1 	\quad 	107 $\pm$ 3	& 4.6 $\pm$ 1.3 	\quad	206 $\pm$ 9	& 1.8 $\pm$ 0.1 	\quad	167 $\pm$ 4	\\
1ES1959+650 I   \textsuperscript{5} & 8 $\pm$ 2 	\quad	123 $\pm$ 8	& 4.5 $\pm$ 0.2 	\quad	159 $\pm$ 1	& -- 			\quad	--			\\
1ES1959+650 II  \textsuperscript{5} & $<$5 	\quad	--			& 4.7 $\pm$ 0.6 	\quad	151 $\pm$ 19	& $<$1.6  		\quad	--			\\
%1ES1959+650 III \textsuperscript{6} & 9 $\pm$ 2 	\quad	53 $\pm$ 5	& 4.5 $\pm$ 0.7 	\quad	152 $\pm$ 6	& 1.3 $\pm$ 0.4 	\quad	140 $\pm$ 9	\\
%1ES1959+650 IV  \textsuperscript{6} & 12 $\pm$ 1 	\quad	20 $\pm$ 2	& 5  $\pm$ 1 		\quad	158 $\pm$ 14	& 2.5 $\pm$ 1 	\quad	149 $\pm$ 11	\\
PG1553+113    	\textsuperscript{6} & 10 $\pm$ 2 	\quad	86 $\pm$ 8	& 4.2 $\pm$ 0.5 	\quad	Rotation 	& 2.6 $\pm$ 0.7   \quad	133 $\pm$ 7 	\\
1ES0229+200  	\textsuperscript{7} & 18 $\pm$ 3 	\quad	25 $\pm$ 5	& 3.2 $\pm$ 0.7 	\quad	$-$5 $\pm$ 9	& $<$7 		\quad	--			\\
% PKS2155-304 	\textsuperscript{8} & 23 $\pm$ 2 	\quad	128 $\pm$ 2	& 4.0 $\pm$ 0.3 	\quad	Rotation 	& 1.7 $\pm$ 0.4 	\quad	113 $\pm$ 6	\\
\bottomrule
\end{tabularx}
\noindent{\footnotesize{\textsuperscript{1--7} Results compiled from the following references: \cite{2022Natur.611..677L, 2022ApJ...938L...7D, DiGesu2023, 2024A&A...681A..12K, 2024arXiv240104420E, 2023ApJ...953L..28M, 2023ApJ...959...61E}; \textsuperscript{a} median polarization properties during the IXPE observation. Especially for optical and IR polarization, only corrected polarization values, accounting for the dilution of polarization by unpolarized starlight from the host galaxy, were considered for calculation; \textsuperscript{b} at the lowest radio frequency (4.85 GHz).}}
%MDPI: Please confirm whether there is a superscript ‘8’  and ‘b’ in the table. please add them in table.
\end{table}

\vspace{-12pt} 
\begin{table}[H]
\caption{X-ray spectral properties of HSPs during \ixpe observation. \label{tab:hsp}}
\begin{tabularx}{\textwidth}{lccl}
\toprule
\textbf{Source} & \textbf{Photon Index} & \textbf{Flux\boldmath{$_{2-8 \;{\rm keV}}$}} & \textbf{Telescopes} \\
 & & {\footnotesize (\boldmath{$\times 10^{-11}$} \ergsc)} & \\
 \midrule
Mrk~501 I   	\textsuperscript{1} & 2.27 $\pm$  0.01 			& 10.0 $\pm$  0.5 		& {\footnotesize \ixpe + \swift +\nustar} \\
Mrk~501 II  	\textsuperscript{1} & 2.05 $\pm$  0.02 			& 21.0 $\pm$  0.6 		& {\footnotesize \ixpe + \swift }\\
Mrk~421 I   	\textsuperscript{2} & 2.97 $\pm$  0.01 			& 8.67 $\pm$  0.03 	& {\footnotesize \ixpe + \xmm +\nustar} \\
Mrk~421 II  	\textsuperscript{3} & 2.32  $\pm$  0.01 \textsuperscript{a}  	& 15.7 $\pm$  0.1 		& {\footnotesize \swift} \\
Mrk~421 III 	\textsuperscript{3} & 2.13  $\pm$  0.01 \textsuperscript{a} 	& 30.2 $\pm$  0.2 		& {\footnotesize \swift} \\
Mrk~421 IV  	\textsuperscript{4} & 2.82 $\pm$  0.01		 		& 22.5 $\pm$  0.4 		& {\footnotesize \ixpe + \xmm}\\
1ES1959+650 I   \textsuperscript{5} & 2.50 $\pm$  0.02  			& 12.4 $\pm$  0.2 		& {\footnotesize \ixpe + \xmm}\\
1ES1959+650 II  \textsuperscript{5} & 2.29 $\pm$  0.01  			& 14.7 $\pm$  0.1 		& {\footnotesize \ixpe + \xmm}\\
%1ES1959+650 III \textsuperscript{6} & 2.50 $\pm$  0.01 			& 8.61 $\pm$  0.04 	& {\footnotesize \ixpe + \xmm + \swift + \nustar}\\
%1ES1959+650 IV  \textsuperscript{6} & 2.98 $\pm$  0.02 			& 31.4 $\pm$  0.6 		& {\footnotesize \ixpe + \swift}\\
PG1553+113    	\textsuperscript{6} & 2.49 $\pm$  0.01 			& 2.55 $\pm$  0.03 	& {\footnotesize \ixpe + \xmm}\\
1ES0229+200  	\textsuperscript{7} & 1.82 $\pm$  0.01    			& 0.84 $\pm$  --		& {\footnotesize \ixpe + \xmm}\\  %MDPI:  Please confirm whether the data here is complete
% PKS2155-304 	\textsuperscript{8} & 2.71 $\pm$  0.02 			& 2.63 $\pm$  0.01 	& {\footnotesize \ixpe + \xmm}\\
\bottomrule
\end{tabularx}
\noindent{\footnotesize{\textsuperscript{1--7} Results compiled from the following references: \cite{2022Natur.611..677L, 2022ApJ...938L...7D, DiGesu2023, 2024A&A...681A..12K, 2024arXiv240104420E, 2023ApJ...953L..28M, 2023ApJ...959...61E}; \textsuperscript{a} average value between variation over time.}}
%MDPI: Please confirm whether there is a superscript 8 in the table
\end{table}

%MDPI:  We change the date into the format of ``day month year", please confirm
Mrk~501 was the first observed HSP blazar. This target was observed at two different epochs (Mrk~501 I: 8--10 March 2022 and Mrk~501 II: 26--28 March 2022). The time-averaged X-ray polarization in the 2--8 keV range of these observations was consistent within a 3$\sigma$ error range, as shown in Table \ref{tab:xpol_hsp}. There was no significant variability within the duration of the individual \ixpe observations over time and energy, and the X-ray activity was in an average state. The polarization in the optical, IR, and radio bands was significantly lower than in the X-rays by a factor of $\sim$2, indicating strong chromatic behavior (Figure \ref{fig:mrk501_421}). Additionally, \pa\ is aligned with the jet direction, $\psi_{43 {\rm GHz}}=119.7\pm11.8$, obtained from the 43 GHz Very Long Baseline Array (VLBA) observation {\cite{2022Natur.611..677L}}. These multiwavelength polarimetric results suggest, among the various proposed {models}, the ``energy-stratified shock acceleration model'' in which particles accelerated by shocks experience increasing turbulence as they move away from the accelerating site, resulting in a decrease in the polarization degree. {Moreover, shocks can push the particles far away, causing a decrease in \pdx. In addition, the energy-stratified shock acceleration model predicts \pax~along the jet axis. Thereby, the different polarization angles at different wavelengths suggest a scenario where particles emitting at different energies experience different magnetic field conditions.}

Mrk~421 has been observed four times  (Mrk~421 I: 4--6 May 2022; Mrk~421 II: 4--6 June 2022; Mrk~421 III: 7--9 June 2022; and Mrk~421 IV: 6--8 December 2022). Mrk~421 I exhibited a slightly higher \pdx\ of $15\%$ compared to Mrk~501 {within a statistical difference of less than $2\sigma$} and a \pdx/\pdo\ ratio of $\sim$5, indicating a stronger chromatic behavior relative to Mrk~501. These results {are consistent with} the energy-stratified shock acceleration model {based on the \pdx/\pdo\ , where a smaller value was predicted in magnetic reconnection, and the TEMZ models}. Subsequent observations of Mrk~421 II and III did not reveal significant polarization information in the time-averaged polarization during the observation periods. However, the time-resolved results, as shown in Figure \ref{fig:mrk501_421}, revealed a smooth rotation of \pax\ by more than 360$^\circ$ between two consecutive observations (Mrk~421 II: from $9\pm1^\circ$ to $130\pm11^\circ$ and Mrk~421 III: from $222\pm7^\circ$ to $360\pm20^\circ$). The rotation rates of \pax\ were measured as 80 $\pm$ 9$^\circ$ per day and 91 $\pm$ 8$^\circ$ per day for the two observations, respectively, while the polarization degrees remained consistent within uncertainty as \pdx$= 10\pm1\%$. During the observed period, the X-ray flux increased by approximately a factor of two between the two observations and then decreased to a level similar to that at the beginning of the observation, and the spectrum properties showed a decrease in the photon index and an increase in the hardness ratio (i.e., the spectrum became harder and flatter), {although we cannot be sure whether this is directly related to the rotation in the polarization angle}. The multiwavelength polarimetry results showed chromatic trends conistent with previous findings. Therefore, the results of Mrk~421 II and III suggest a geometric structure where the energy-stratified shock propagates linearly or radially along the helical magnetic field within the jet \cite{DiGesu2023}. Additionally, the Mrk~421 IV showed similar characteristics to those of Mrk~421 I in terms of \pdx, but \pax\ exhibited {a different value compared to previous observations (Mrk~421 I, II, and III) and varied across all the observations.}. In particular, Mrk~421 IV confirmed temporal episodic variability in polarization during this observation period. In particular, Ref.~\cite{2024A&A...681A..12K} showed a possible correlation between the rotation of \pax\ and the observation of a highly polarized knot emerging from the jet core and moving away from the ``core'' based on the 43 GHz VLBA observations performed during the observation periods of Mrk~421 II, III, and IV. In addition, multiwavelength polarization monitoring of Mrk~421 I, II, III, and IV showed a rotation of \pa\ in the optical and radio bands over a longer period compared to the X-ray band, with a direction opposite to the rotation of \pax. This result suggests the possibility that emission regions at longer wavelengths (millimeter, infrared, and optical) were separate from, {or only partially coincident with, those of the X-ray emission but also contain a helical magnetic field that is presumably a continuation of the helical field manifested in the X-ray polarization observations.}

\begin{figure}[H]

        \includegraphics[width=0.49\textwidth]{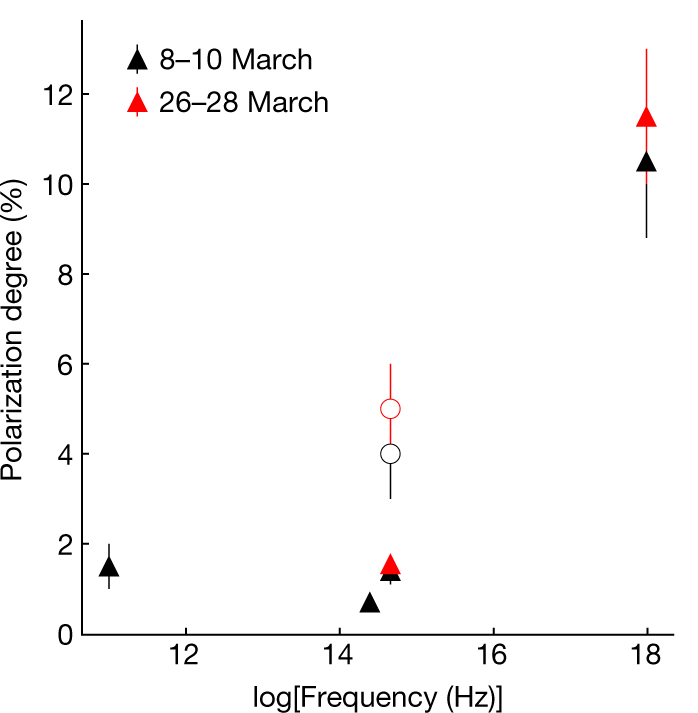}
        \includegraphics[width=0.49\textwidth]{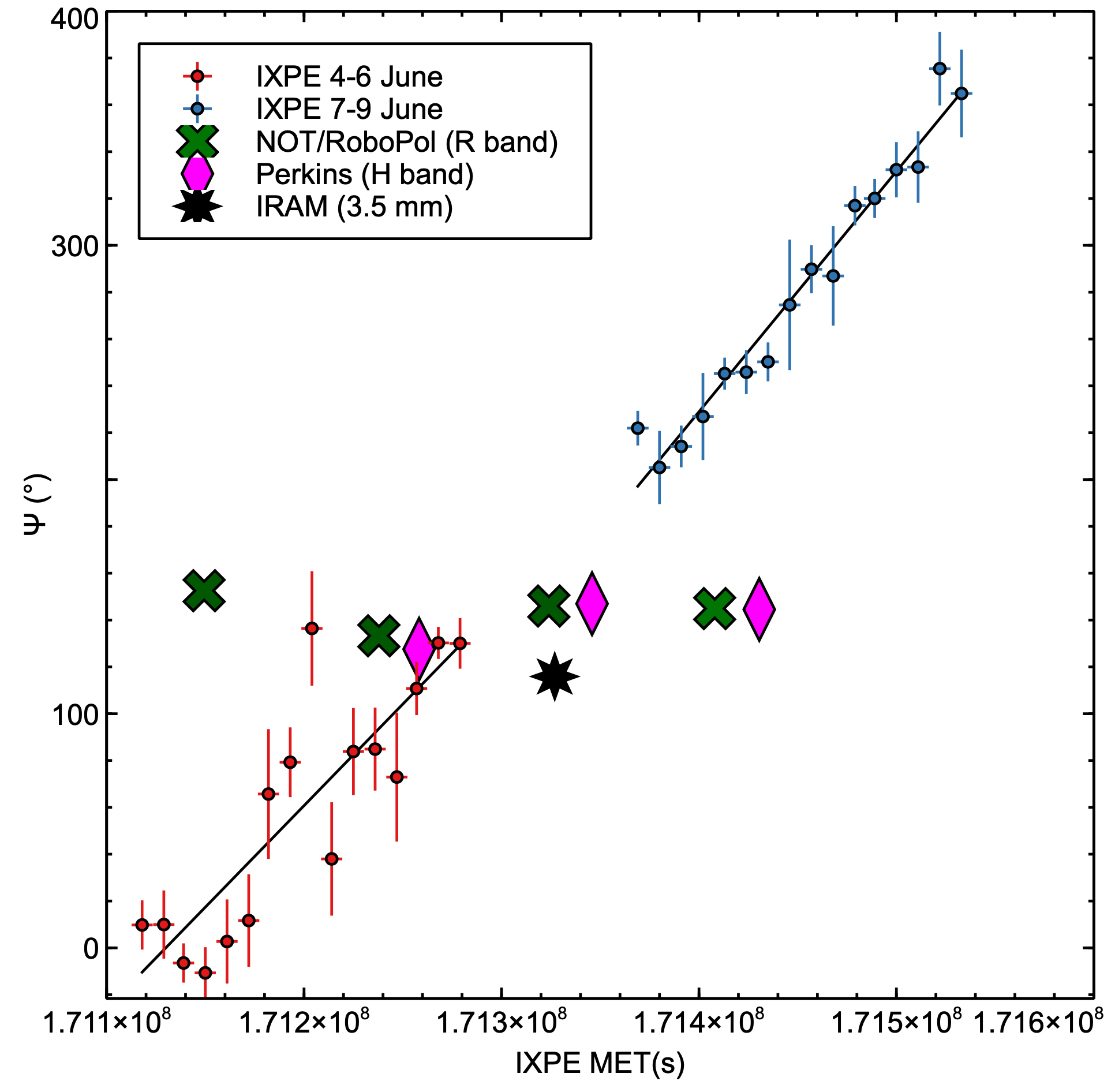}
%        \caption{Rotated Mrk~421}
%        \label{fig:mrk421_rot}
 
    \caption{\ixpe observation results of Mrk~501 and Mrk~421. (\textbf{Left}) Multiwavelength polarization degree of Mrk~501 from radio rays to X-rays. {The black symbols represent observations conducted between March 8th and 10th, while the red symbols represent observations from March 26th to 28th. The open symbols indicate the intrinsic optical polarization degree corrected for the host galaxy.} The figure was reproduced from Figure 3 in \cite{2022Natur.611..677L}. (\textbf{Right}) X-ray polarization angle rotation in Mrk~421. The symbols identify multiwavelength polarimetry measurements obtained from telescopes as labeled. The figure was reproduced from Figure 2 in \cite{DiGesu2023}.}
    \label{fig:mrk501_421}
\end{figure}

\textls[-15]{In addition, the observations of 1ES1959+650 \cite{2024arXiv240104420E}, PG1553+113 \cite{2023ApJ...953L..28M}, and 1ES0229+200~\cite{2023ApJ...959...61E}} exhibited significant chromatic behavior on \pd\ across multiwavelengths, {consistent with} the energy-stratified shock acceleration model. Furthermore, concerning energy stratification and cooling inside the shock, the \pdx/\pdo ratio from all the \ixpe HSP observations has typically been found to be around $\sim$2--6, with a trend between the \pdx/\pdo\ ratio and \pdx\ , as shown in the left panel of Figure \ref{fig:global_hsp}. {Consequently, as \pdx/\pdo\ increases, X-ray emission occurs in smaller volumes due to a relatively shorter cooling timescale, indicating well-ordered magnetic fields closer to the shock front and increased turbulence in the downstream direction \cite{2023ApJ...948L..25P}}. Moreover, the comparison between the \pdx/\pdo\ ratio and the time-averaged photon index in the right panel of Figure \ref{fig:global_hsp} suggests a possibility of relatively higher \pdx/\pdo\ ratios when the spectrum is softer and steeper, but excluding the observations of 1ES0229+200. Especially in the case of 1ES0229+200, an $E_{pivot}$ scale factor of 1 was used in the log parabolic model instead of the common 5 used in other observations~{\cite{2024A&A...681A..12K, DiGesu2023, 2022Natur.611..677L}}. Therefore, it is possible that this factor may have influenced the observed results compared to other observations, and this could be improved through analysis under similar conditions in the future. Additionally, this characteristic can be interpreted based on the same physical interpretation as when the X-ray emission originates further away from the peak frequency on the spectrum, indicating that the emission is from well-ordered magnetic fields in a smaller volume with a shorter cooling timescale closer to the shock~front.

Regarding the variability in polarization properties, smooth rotation on \pax\ and transient \pax\ variation have been observed only in Mrk~421. Hence, it is speculated that the polarization variability observed exclusively in specific targets may be linked to the inherent physical properties of jets. However, it is imperative to scrutinize statistically significant results through concurrent observations based on a larger sample of targets in future studies. Finally, concerning the variability identified through the multiwavelength polarimetry observations of Mrk~421 and an orphan polarization rotation exhibited by PG1553+113 supports the scenario that the X-ray emission region is separate from that at longer wavelengths.

\begin{figure}[H]
  
         \includegraphics[width=\textwidth]{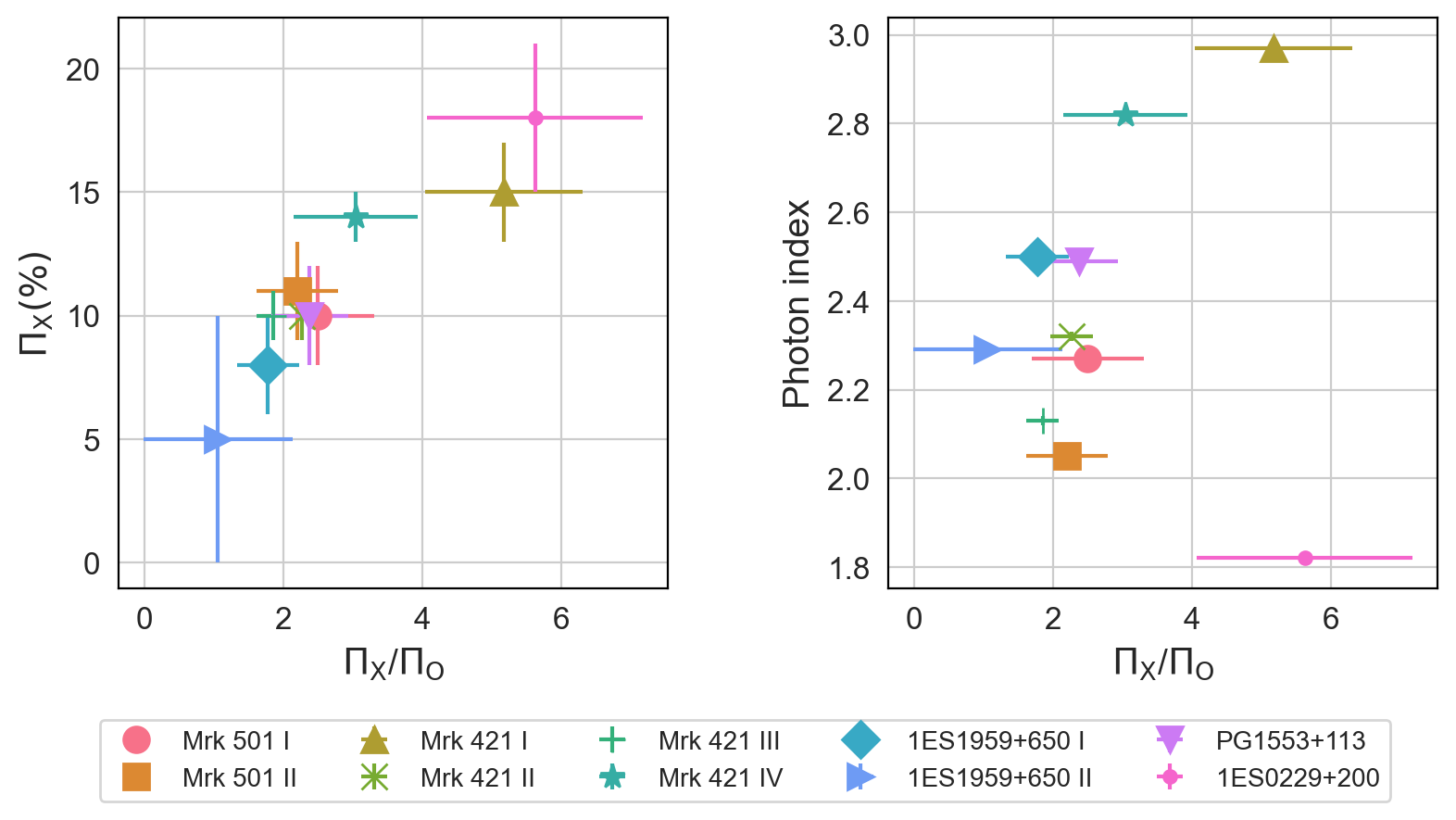}
\caption{Physical properties of energy-stratified shock emissions from HSP blazars. (\textbf{Left}) X-ray polarization vs. the \pdx/\pdo ratio; (\textbf{Right}) X-ray power law photon index vs. the \pdx/\pdo\ ratio of HSP blazars observed by \ixpe. Each colored point indicates different \ixpe observations labeled in the~legend.}\label{fig:global_hsp}
\end{figure}

%\begin{figure*}[t!]
%         \centering
%         \includegraphics[width=\textwidth]{figure/hsp_del_psi.png}
%\caption{The angle difference between \pax\ and the jet direction observed by VLBA \cite{2022ApJS..260...12W}. Each color represents the $\Delta\Psi$ between the jet direction and \pax\ for a different \ixpe observation, labeled identically to Figure \ref{fig:global_hsp}. The Red dashed line indicates the jet direction observed by VLBA, and the shaded area shows the error range of measurement. For \pa\ exceeding the $-90^\circ$ to $+90^\circ$ range, the value was adjusted by $\pm n\pi$. In particular, \pax\ of Mrk~421 II and III is shown as angle range sweep by rotation.}\label{fig:del_psi}
%\end{figure*}

%\begin{table}[H]
%\centering
%\caption{Physical parameter of Jets form VLBA observation in 43 GHz\label{tab:hsp_jet}}
%\begin{tabularx}{0.75\textwidth}{lc}
%\toprule
%\textbf{Source} & \textbf{EVPA of the jet}\\
%\midrule
%Mrk~501     &  119.7$\pm$11.8    & 53.8 $\pm$ 11.8 \\
%Mrk~421     &  -14.4$\pm$14.2    & 27.8 $\pm$ 14.2 \\
%1ES1959+650 & 127.9$\pm$12.9    & 38.2 $\pm$ 12.9 \\
%
%\bottomrule
%\noindent{\footnotesize{Results compiled from \cite{2022ApJS..260...12W}}}\\
%\end{tabularx}
%\end{table}

\subsection{\ixpe Observations of Intermediate and Low Synchrotron Peak Blazars}

In the case of ISP and LSP blazars, \ixpe observations have been carried out for BL Lac~\cite{2023ApJ...942L..10M, 2023ApJ...948L..25P}, 3C~273, 3C~279, 3C~454.3, and S5~0716+714 \cite{2023arXiv231011510M}. BL Lac is typically classified as an LSP or ISP, where the Compton hump dominates the soft X-ray band. The X-ray flux and photon index of this target are typically $<2\times10^{-11}$\ergs and $ < 2$, respectively. BL Lac has been observed by the IXPE three times: during an average flux state (BL Lac I and II;~\citep{2023ApJ...953L..28M}) and during an outburst (BL Lac III; \citep{2023ApJ...948L..25P}). From BL Lac I and II, the upper limits of \pdx~$\leq~14.2\%$ and \pdx~$\leq~12.6\%$ were estimated, respectively, and the median optical polarization values corresponding to \pdo~=~6.8$\%$,  \pao~=~107$^\circ$ (during BL Lac I), and \pdo~=~14.2$\%$ and \pao~=~42$^\circ$ (during BL Lac II) were measured from contemporaneous multiwavelength polarimetry observations. These results were interpreted as follows: (1) In the leptonic scenario, considering that emission originating from SSC dominates over EC emission in the \textit{IXPE} energy band according to SED modeling, approximately$\sim$3$\%$ of \pdx\ is predicted from given polarization in radio band, \pdradi (BL Lac I$\sim$4$\%$ and BL Lac II$\sim$$6\%$) \cite{2019ApJ...885...76P}. Therefore, the measured X-ray upper limit is consistent with the leptonic scenario. (2) In the hadronic scenario, \pdx\ similar to or higher than \pdo\ is predicted, but \pdo\ was observed to be high. Therefore, the hadronic scenario is disfavored. In conclusion, this disfavors a significant contribution of proton synchrotron radiation to the X-ray emission at these epochs. Instead, it supports a leptonic origin for the X-ray emission in BL Lac.
%MDPI:  We have moved these contents out of the references command, please confirm.

Furthermore, in the case of BL Lac III, a distinct synchrotron contribution was observed alongside the outburst, with $F_{2-8 {\rm keV}}~=~2.77\pm0.21 \times 10^{-11}$\ergs~and a photon index of 2.10 $\pm$ 0.09. Subsequently, Ref.~\cite{2023ApJ...948L..25P} investigated the flux and polarization contributions of the synchrotron emissions across both the soft-energy range (2--4 keV) and the entire \textit{IXPE} band (2--8 keV) using two power law components. Through time-resolved polarization analysis, dividing the total observation period into three equal sub-periods, a significant X-ray polarization, exceeding $99\%$ significance, with \pdx~=~21.7$^{+5.6}_{-7.9}$ and \pax~=~$-$28.7 $\pm$ 8.7, was detected only in the first sub-period (upper limits were observed for the remaining sub-periods). This was accompanied by a decrease in the flux contribution of the synchrotron emission component. {Moreover, the \pdx\ value was higher than the concurrently measured \pdo~=~13.1 $\pm$ 2.4$\%$.} These results were explained by the influence of turbulence in the magnetic field, closer to the particle acceleration site in a smaller volume with well-ordered magnetic fields, akin to the characteristics of synchrotron emission observed in HSPs, leading to the higher \pdx\ observed in the high-energy regime. Consequently, BL Lac III was concluded to probe the high-frequency end of the synchrotron component, which diminished during the observation.
%MDPI: The Ref. number is not allowed to be used as the subject in sentences of the main text, advise to add “Ref.” or “author names” before the citation, We corrected them. Please confirm

For the remaining LSP and ISP sources (3C~273, 3C~279, 3C~454.3, and S5~0716+714), only upper limits could be estimated from \textit{IXPE} observations in the 2--8 keV band (3C~273: \pdx ~<~$9.0\%$, 3C~279: \pdx~<~$12.7\%$, 3C~454.3: \pdx~<~$28.0\%$, and S5~0716+714: \pdx~<~$26.0\%$). These observational results, when compared with polarization in the optical and radio domains, suggest that \pdx~<~$\sim10\%$ and is likely to be less than or comparable to contemporaneous \pdo. {Therefore, although hadronic models are not completely eliminated, these X-ray polarization measurements seem to favor the leptonic process, wherein photons are upscattered by relativistic electrons in the jets of blazars. Lastly}, Ref.~\cite{2023arXiv231011510M} emphasizes that further \ixpe observations can provide additional insights into the X-ray emission mechanism in LSP and ISP blazars.

\section{X-ray Polarization from Hot Coronae}\label{sec:xpol_corona}

\subsection{Theoretical Expectations}

The geometry and size of the X-ray corona serve as indicators for exploring its physical origin. The corona could manifest as a slab-like or wedge-like structure originating from the accretion disk, e.g., \citep{1993MNRAS.261..346H, 2023ApJ...949L..10P}, and it could assume a conical shape if it forms the base of a jet, e.g., \citep{1991ApJ...383L...7H, 1997A&A...326...87H, 2005ApJ...635.1203M}, or a failed jet, e.g.,  \citep{2004A&A...413..535G}. Often, it has been assumed for simplicity as a point- or sphere-like configuration located above and below the black hole rotation axis{, called lamppost geometry}, e.g., \citep{1996MNRAS.282L..53M, 2004MNRAS.349.1435M, 2012MNRAS.424.1284W}. 

The X-ray polarization from the corona is expected to depend on its size, geometry, and position in the system. Recent simulation results using the Monte Carlo Comptonization code (MONK) \cite{2019ApJ...875..148Z, 2022MNRAS.510.3674U} yield the following polarizations for different corona geometries: (1) the slab geometry is predicted to {exhibit} a maximum polarization, \pdx\, {up to} $12\%$, with \pax\ oriented {parallel  to the disk axis}, i.e., parallel to the direction of the jet{; (2) conical geometry is expected to result in }\pax\ {parallel to the major axis of the projection of the disk on the sky} with \pdx\ up to $7\%$; (3) {in the case of spherical lamppost geometry}, \pdx\ of $1-3\%$ and {the same }\pax\ {as conical geometry are predicted} (see Figure \ref{fig:corona_pol}).

\begin{figure}[H]
      
         \includegraphics[width=0.95\textwidth]{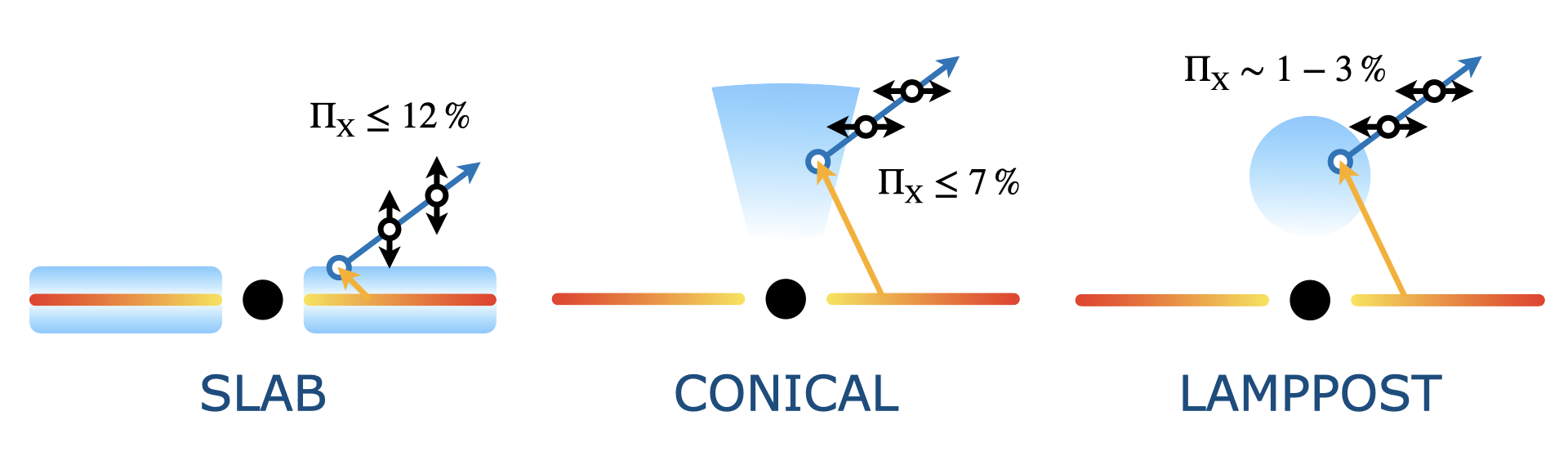}
\caption{Schematic view of different X-ray corona geometries: slab, conical, and lamppost. The bluish and reddish areas represent the corona and disk emission regions, respectively. Blue and orange arrows indicate upscattered X-ray and UV/optical disk photons, respectively. The black arrow represents the predicted polarization properties (degree in number and angle in direction of the arrow), estimated from \cite{2022MNRAS.510.3674U}.}\label{fig:corona_pol}
\end{figure}

\subsection{\ixpe Observations of Unobscured Radio-Quiet AGN}

Since the beginning of IXPE operation, three unobscured radio-quiet Seyfert 1 galaxies have been observed (NGC~4151: \cite{2023MNRAS.523.4468G}, IC~4329A: \cite{2023MNRAS.525.5437I}, and MCG-05-23-16: \cite{2023MNRAS.525.4735T, 2022MNRAS.516.5907M}). {The X-ray polarization analysis for RQAGN was conducted using both model-independent and spectropolarimetric analysis methods. These methods were employed to examine the time-averaged polarization and the variability in polarization over time and energy.} In particular, the relatively more complex spectral features of RQAGN were examined based on simultaneous observations from \xmm\ and \nustar, which cover a wider energy range with higher spectral resolutions. This allows for a more detailed spectropolarimetric analysis, resulting in the measurement of X-ray polarization characteristics for each emission component. Finally, the polarization properties measured by the \ixpe\ were used to constrain the geometric characteristics of the corona by comparing the expected X-ray polarization results predicted under the different geometrical conditions of the corona with the MONK simulation results.

First, NGC~4151 exhibited X-ray polarization in the 2--8 keV energy range, with \pdx~=~4.9 $\pm$ 1.1$\%$ and \pax~=~86 $\pm$ 7$^\circ$ (as shown in the left panel of Figure \ref{fig:ngc4151}), corresponding to $\sim$4.4$\sigma$ significance, as determined by a model-independent polarization analysis. Moreover, NGC~4151 revealed statistically significant variations in the energy-resolved polarization analysis; i.e., we found \pdx~=~4.3 $\pm$ 1.6$\%$ and \pax~=~42 $\pm$ 11$^\circ$ in the soft-energy range of 2--3.5 keV, deviating from the measurements in other energy ranges. Additionally, spectropolarimetric analysis (middle panel in Figure \ref{fig:ngc4151}) takes into account the polarization dilution effect due to the soft-energy regime and sets the polarization angles of the primary emission and reflection emission components perpendicular to each other. As a result, for the primary emission component, \pdx~=~7.7 $\pm$ 1.5$\%$ and \pax~=~87 $\pm$ 6$^\circ$ were derived, while, for the reflection component, an upper limit of \pdx$<27\%$ was obtained. {Additionally, NGC~4151 exhibits a radio jet feature in VLBI {observations} \cite{2017MNRAS.472.3842W}, and the measured \pax\ shows that it aligns with the direction of the radio jet emission, $\sim$83$^\circ$ \cite{1986MNRAS.218..775H, 1998ApJ...496..196U}}. This indicates that the corona emission originates from the equatorial plane. These polarization results ruled out lamppost and conical geometry scenarios among the proposed coronal geometries. Furthermore, a comparison with the MONK simulations (right panel in Figure \ref{fig:ngc4151}) confirmed that the \ixpe\ observational results agree well with the predicted polarization results from the slab or wedge geometry. Moreover, the observed \pax\ in the soft-energy band is similar to that of the extended narrow-line region (NLR), with a confirmed orientation by HST \cite{1993ApJ...417...82E, 2003AAS...203.5601D}, supported by a Chandra observation \cite{2011ApJ...742...23W} value of $45\pm5^\circ$, suggesting that a new emission component may  {influence} the X-ray polarization, which is to be confirmed through additional observations.

\begin{figure}[H]
    \centering       
        \begin{subfigure}[b]{0.24\textwidth}
        \includegraphics[width=\textwidth]{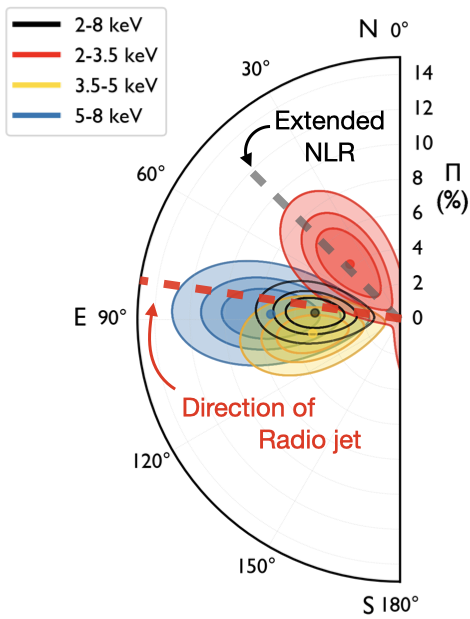}
    \end{subfigure} 
    \begin{subfigure}[b]{0.38\textwidth}
        \includegraphics[width=\textwidth]{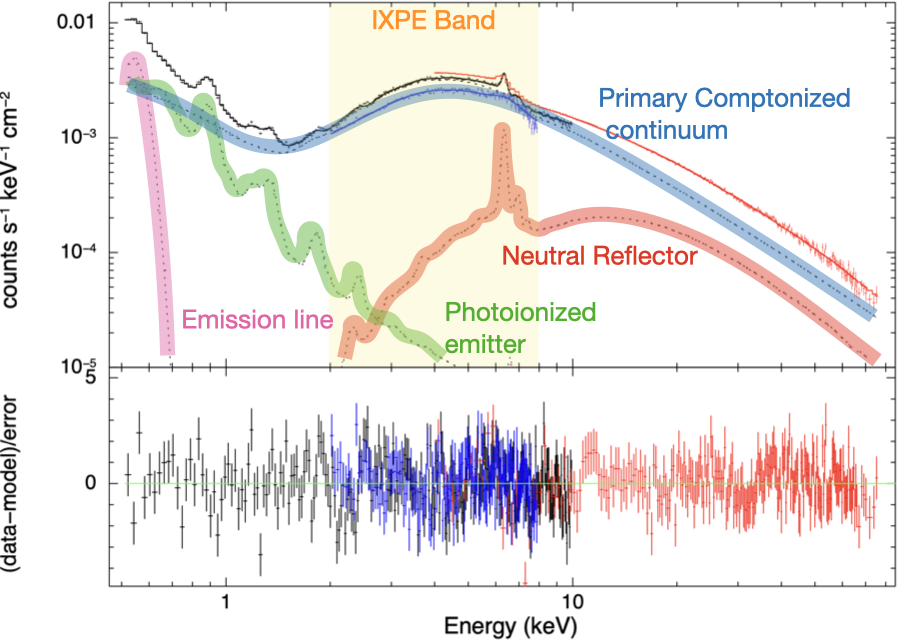}
    \end{subfigure}
    \begin{subfigure}[b]{0.35\textwidth}
        \includegraphics[width=\textwidth]{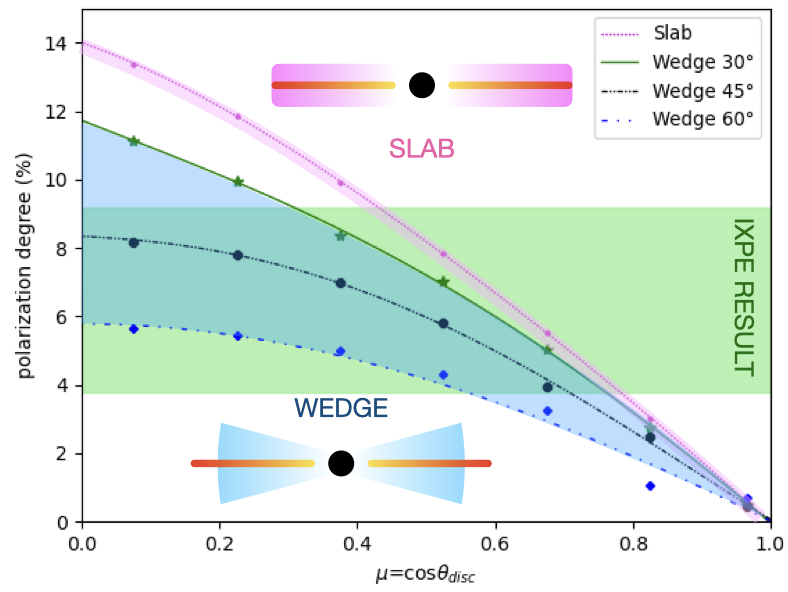}
    \end{subfigure}
    \caption{\ixpe observation result of NGC~4151. (\textbf{Left}) Polarization contours ($68\%$, $90\%$, and $99\%$ detection significance levels); (\textbf{middle}) X-ray spectrum analysis result; (\textbf{right}) comparison of \ixpe measurements and expected X-ray polarization in slab and wedge geometry coronae calculated from MONK simulations. The figure was {reproduced with kind permission from Oxford University Press and the Royal Astronomical Society from} \cite{2023MNRAS.523.4468G}.\label{fig:ngc4151}}
    
\end{figure}

Next, MCG-05-23-16 underwent two observations (MCG-05-23-16 I: May 2022 and MCG-05-23-16 II: November 2022), yielding upper limits of \pdx~$\leq 4.7\%$ and \pdx~$\leq 3.3\%$, respectively. The integrated analysis of MCG-05-23-16 I and II resulted in an upper limit of \pdx~$\leq 3.2\%$, and Figure \ref{fig:mcg_5_23_16} illustrates the comparison of the X-ray polarization contour plot with the predicted polarization characteristics from the MONK simulations of the proposed corona geometries. Thus, despite \pa\ being estimated at a $\sim$$1 \sigma$ confidence level, considering the direction of the NLR estimated by the HST [O III] emission observations to be $\sim$$40^\circ$ \cite{2000ApJS..128..139F}, and taking into account the perpendicular orientation of the NLR and the accretion disk, the results hint at a coronal geometry in the form of a slab or wedge rather than a lamppost or conical shape. Additionally, the disk inclination estimated in \cite{2023MNRAS.526.3540S} of $\sim$30--50$^\circ$ supports the results of a coronal structure like a slab or wedge.

\begin{figure}[H]
         \includegraphics[width=\textwidth]{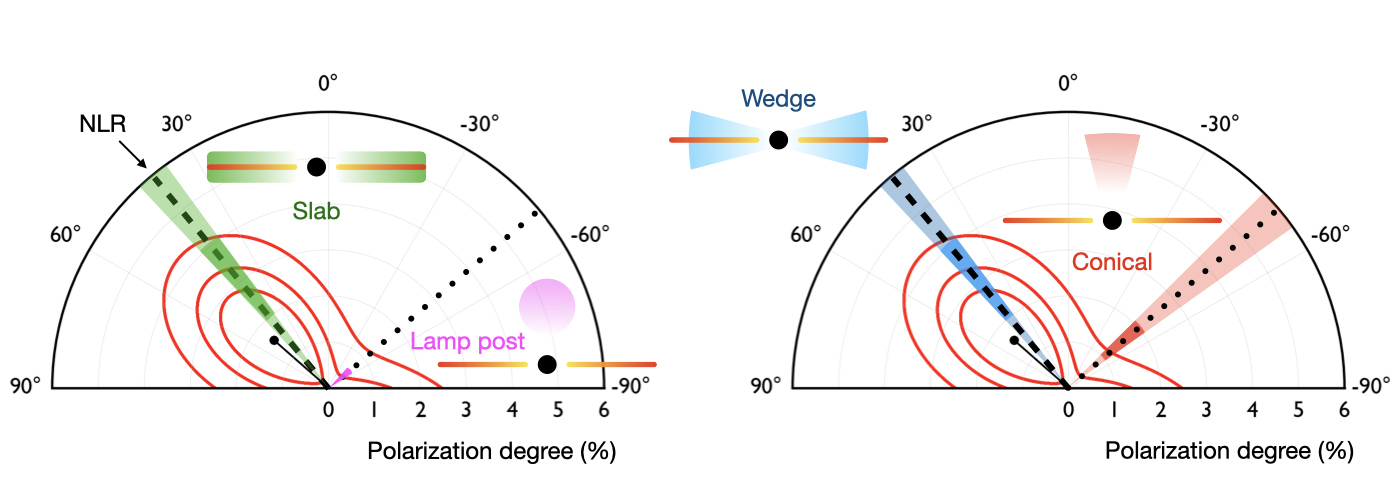}
\caption{X-ray polarization contours of the integrated analysis of MCG-05-23-16 I and II (red; $68\%$, $90\%$, and $99\%$ detection significance levels). The black dashed line indicates the direction of the NLR, while the dotted line denotes the perpendicular direction, implying an angle parallel to the disk. Each colored area represents the expected polarization properties calculated by MONK, with relatively saturated regions indicating the expected degree of polarizaton for inclinations in the range of approximately 30--50$^\circ$. The figure was {reproduced with kind permission from Oxford University Press and the Royal Astronomical Society from} Figure 8 in \cite{2023MNRAS.525.4735T}.}\label{fig:mcg_5_23_16}
%MDPI: Please change the hyphen (-) into a minus sign (−), e.g., “-1” should be “−1”. 
\end{figure}

Finally, IC~4329A has a detected X-ray polarization of the hot corona of $3.3\pm1.1\%$ for \pdx\ and $78\pm10^\circ$ for \pax\ in the 2--8 keV band, utilizing simultaneous observational data from \ixpe, \xmm, and \nustar\ for the spectropolarimetric analysis. For the spectral model, a Comptonization and a reflection component have been applied. The \pax\ was roughly aligned with the position angle of the extended radio emission at $\sim$90$^\circ$, e.g.,~\citep{1987MNRAS.228..521U}. This result, consistent with NGC~4151 and MCG-05-23-16, supports the geometric properties of a radially extended corona within the plane of the disk. Additionally, in~\cite{2023MNRAS.525.5437I}, through a more detailed examination using MONK simulations, a test of the slab and wedge geometry corona system has been conducted. This tentatively suggests, at a 90\% confidence level, a lower limit for \pdx, favoring asymmetric configurations possibly characterized by outflowing coronal geometries extended within the disk plane, which would generate highly polarized emissions aligned with the jet.

%%%%%%%%%%%%%%%%%%%%%%%%%%%%%%%%%%%%%%%%%%
\section{Summary}

The \ixpe mission has pioneered X-ray polarimetry, achieving the first-ever detections of X-ray polarization from {AGN}. The diverse physical characteristics of {AGN} observed in the X-ray range, particularly in the RLAGN population represented by blazars and the RQAGN population represented by unobscured RQAGN, have provided crucial insights into estimating the particle acceleration processes within relativistic jets, the physical and geometrical properties of magnetic fields, and the intrinsic geometry of hot coronae. The \ixpe observations of blazars, notably the HSP subclass starting with the first observation of Mrk~501, {are consistent with the possibility that the synchrotron emission within the jet originates from energy-stratified shocks}. Subsequent observations of Mrk~421 have further supported this scenario and indicated the presence of helical magnetic fields by discovering the smooth rotation of the polarization angle for the first time in the X-ray range. Additionally, other HSP observations, including 1ES1959+650, PG1553+113, and 1ES0229+200, showed values of the \pdx/\pdo\ ratio around $\sim$2--6, {supporting} the conclusions drawn from Mrk501 and Mrk421. Furthermore, extensive comparisons of all the HSP observations have led to a physical interpretation suggesting that synchrotron jet emission at high frequencies takes place in environments with smaller volumes of well-ordered magnetic fields close to shock fronts. The polarization information at longer wavelengths is diluted by magnetic field turbulence. This interpretation could gain further solidity through additional \ixpe observations focusing on spectral properties and polarization angles. Furthermore, temporal variations in polarization properties have been solidly observed only in Mrk~421 to date. Hence, the characteristic of significant polarization variability observed only in specific targets should be meticulously investigated in future studies, considering additional parameters such as the physical characteristics of the jet and utilizing more extensive observational data. Lastly, the individual variations in polarization observed across multiple wavelengths imply spatial segregation between X-ray emissions and emissions at other wavelengths.

\ixpe observations of other subclasses of blazars, including LSPs and ISPs, were carried out to explore the particle acceleration mechanisms of the Inverse Compton emission component situated in the jet's higher-energy band. The observations of BL Lac objects confirmed that the polarization characteristics predicted by the leptonic-based SSC acceleration mechanisms {better} explain the \ixpe observations than hadronic scenarios. However, the relatively low X-ray flux compared to HSPs poses challenges in achieving statistically significant observational results for many targets. Thus, further exploration through observations during higher-source activity states such as future outbursts or longer exposure times should be considered.

The \ixpe observations of {RQAGN} explore the geometry of the X-ray corona centered around the unobscured central engine. Observations of NGC4151, MCG-05-23-16, and IC4329A with \ixpe detected polarization within 3--5$\%$ from the coronal  emission, {aligned with the inferred jet direction}. These results suggest that the corona, instead of being located on the black hole axis as expected if it is the base or possibly failed jet, is better explained by a slab-like or wedge-like shape, believed to originate from Comptonization around the accretion disk, such as a  disk outflow. The \ixpe findings were further validated through a comparison with the predicted X-ray polarization characteristics calculated via MONK simulations based on various geometrical structures. Furthermore, observations of NGC~4151 suggested the possibility of other emission components in the soft-energy band influencing the X-ray polarization, a characteristic that will be further explored in depth through future additional \ixpe observations. Finally, future \ixpe observations aimed at probing the different states of the corona emission process and reflection emission in the accretion disk will provide additional evidence to expand our understanding of the physical emission processes occurring inside {AGN}.

%%%%%%%%%%%%%%%%%%%%%%%%%%%%%%%%%%%%%%%%%%
%\section{Patents}
%
%This section is not mandatory, but may be added if there are patents resulting from the work reported in this manuscript.

%%%%%%%%%%%%%%%%%%%%%%%%%%%%%%%%%%%%%%%%%%
\vspace{6pt}

\vspace{6pt} 
\authorcontributions{Conceptualization, formal analysis, visualization, and writing—original draft preparation, D.E.K.; writing—review and editing, L.D.G., F.M., A.P.M., G.M., P.S., F.T., E.C. and I.D. All authors have read and agreed to the published version of the manuscript.}

\funding{The Imaging X-ray Polarimetry Explorer (IXPE) is a joint US and Italian mission. The US contribution is supported by the National Aeronautics and Space Administration (NASA) and led and managed by its Marshall Space Flight Center (MSFC), with industry partner Ball Aerospace (contract NNM15AA18C). The Italian contribution is supported by the Italian Space Agency (Agenzia Spaziale Italiana, ASI) through contract ASI-OHBI-2017-12-I.0, agreements ASI-INAF-2017-12-H0 and ASI-INFN-2017.13-H0, and its Space Science Data Center (SSDC), and by the Istituto Nazionale di Astrofisica (INAF) and the Istituto Nazionale di Fisica Nucleare (INFN) in Italy. This research used data products provided by the IXPE Team (MSFC, SSDC, INAF, and INFN) and distributed with additional software tools by the High-Energy Astrophysics Science Archive Research Center (HEASARC) at NASA Goddard Space Flight Center (GSFC). The research of APM was partially supported by National Science Foundation grant AST-2108622, NASA Swift grant 80NSSC23K1145, and NASA NuSTAR grant 80NSSC22K0537.} %MDPI: We removed Acknowledgments part since the content in Acknowledgments which should belong to Funding part, Acknowledgments and Funding cannot be the same. (Acknowledgments part is optional)

\dataavailability{{IXPE data are publicly available at} \url{https://heasarc.gsfc.nasa.gov/docs/heasarc/missions/ixpe.html} (accessed on 17 March 2024).} %MDPI: We encourage all authors of articles published in MDPI journals to share their research data. In this section, please provide details regarding where data supporting reported results can be found, including links to publicly archived datasets analyzed or generated during the study. Where no new data were created, or where data is unavailable due to privacy or ethical restrictions, a statement is still required. Suggested Data Availability Statements are available in section “MDPI Research Data Policies” at https://www.mdpi.com/ethics.
%\acknowledgments{\hl{~}} %MDPI: Acknowledgments are optional, please confirm if you need to add content (acknowledgments and funds cannot be exactly the same), or you can delete it directly
\conflictsofinterest{{The authors declare no conflict of interest.}}%MDPI: Declare conflicts of interest or state “The authors declare no conflict of interest.”.

\begin{adjustwidth}{-\extralength}{0cm}
%\printendnotes[custom] % Un-comment to print a list of endnotes

\reftitle{References}

%=====================================
% References, variant A: external bibliography
%=====================================

\PublishersNote{}

\end{adjustwidth}
\end{document}